\documentstyle[preprint,aps]{revtex}	
\begin{document}
\draft
%
%
\preprint{UCD--94--9}
\title{Hadronic $ZZ$, $W^- W^+$, and $W^{\pm}Z$ Production\\
with\\
QCD Corrections and Leptonic Decays}
\author{J.~Ohnemus}
\address{
Department of Physics\\
University of California\\
Davis, CA 95616 USA}
\maketitle
\begin{abstract}
The processes $p\,p\hskip-7pt\hbox{$^{^{(\!-\!)}}$} \rightarrow
V_1 V_2 + X \rightarrow \ell_1 \bar \ell_1 \ell_2 \bar \ell_2 + X$,
where $V_i = W^{\pm}$ or $Z$ and $\ell_i$ denotes a lepton,
are calculated to ${\cal O}(\alpha_s)$.
Total and differential cross sections, with
acceptance cuts imposed on the final
state leptons, are given for the Tevatron
and LHC center of mass energies.
Inclusive and
exclusive $0$-jet and $1$-jet cross sections are given.
The transverse momenta spectra of the leptons are significantly enhanced at
high $p_T^{}$ by the QCD radiative corrections, especially at the
LHC energy.  Invariant mass and angular distributions are scaled up in
magnitude by the QCD radiative corrections, but are little changed in shape.
\end{abstract}
\vskip .5in
\pacs{PACS numbers: 12.38.Bx, 14.80.Er}
\newpage
%
%
\narrowtext

\section{INTRODUCTION}

The production of massive weak boson pairs ($ZZ$, $W^- W^+$, and $W^{\pm}Z$)
is an important physics topic at high energy hadron colliders.
Measurements of these processes are vital for testing the Standard Model (SM)
and probing beyond it.  In particular, these processes are important for
elucidating the electroweak symmetry breaking mechanism and for testing the
triple weak boson coupling \cite{EHLQ}.
Futhermore, these processes are also backgrounds
to new physics signals, for example, new heavy particles such as neutral
and charged Higgs bosons, techni-mesons, extra gauge bosons,
squarks, and gluinos can all decay into weak boson pairs.
It is therefore important to have precise calculations of hadronic weak boson
pair production.  These calculations should include the leptonic decays of the
weak bosons since the weak bosons are identified via their leptonic decay
products.

The dominant production mechanism for weak boson pairs in hadronic collisions
is via the quark-antiquark annihilation subprocess
$q_1 \bar q_2 \to V_1 V_2$.  The cross sections for hadronic
$ZZ$, $W^- W^+$, and $W^{\pm}Z$ production were first calculated in
Refs.~\cite{BROWN:ZZ} and \cite{BROWN:WZ}.
Tree level calculations of $V_1 V_2 + n$~jets have been given
for $n \le 1$ and $n \le 2$ in Refs.~\cite{VVJETONE} and \cite{VVJETTWO},
respectively.
$ZZ$ and $W^- W^+$ can also be produced via the gluon fusion subprocess
$gg \to VV$, which proceeds via a quark box loop
and is of order $\alpha_s^2$ \cite{GGFUSION:ZZ,GLOVER:GG,GGFUSION:WW,GEORGI}.
The cross section for the gluon fusion process is significant at
supercollider energies due to the large gluon luminosity, but it
never dominates the $q \bar q$ annihilation cross section.
Weak boson pairs can also be produced via the vector boson fusion process
in which the incoming quarks radiate two vector bosons which subsequently
scatter off each other \cite{VVFUSION:WW}.
This process is mainly of interest as a source of
Higgs bosons, with the Higgs boson appearing as an $s$-channel resonance.
Away from the Higgs boson peak, the weak-boson fusion production rate is only
a small fraction of the basic $q_1 \bar q_2 \to V_1 V_2$ production rate.
The QCD radiative corrections to hadronic
$ZZ$\cite{ZZ,MELE}, $W^- W^+$\cite{WW,WWFRIX},
and $W^{\pm} Z$\cite{WZ,FRIX}
production have recently been calculated for the case of real weak bosons
in the final state.

In this paper, previous calculations of next-to-leading-order (NLO)
$ZZ$\cite{ZZ}, $W^- W^+$\cite{WW}, and $W^{\pm} Z$\cite{WZ}
production are extended to include the leptonic decays of the
$W$- and/or $Z$-bosons.
Since it is the decay products that are observed in an experiment, the
inclusion of the leptonic decays in the calculation will make it much more
useful for comparing with experimental data.
Futhermore, cuts
can now be applied to the final state leptons, thus allowing one to mimic
the experimental conditions.  The angular distributions of the weak boson's
decay products are especially important because they are effective spin
analyzers for the vector bosons \cite{WILLEN}.
They are also of crucial importance when one
tries to distinguish the various sources of anomalous couplings in the
three-boson vertex \cite{WILLEN}.  The NLO QCD corrections will have a
significant effect on the distributions of the final state leptons,
especially at the LHC center of mass energy.

The calculations presented here include the leptonic decays
of the weak bosons in the narrow width approximation.
In this approximation, non-resonance diagrams ({\it e.g.},
the final state $e\bar e \mu \bar \mu$, which is formed by the production
and decay of $ZZ$, can also be reached by the process
$p\,p \to Z \to e \bar e$ with the subsequent radiation of a leptonically
decaying $Z$ [$Z \to \mu \bar \mu$] from the electron line)
are not necessary to maintain gauge invariance.
The calculations are done by using
the Monte Carlo method for NLO calculations \cite{NLOMC}
in combination with helicity amplitude methods \cite{HELICITY}.
The Monte Carlo method for NLO calculations is very powerful because
any number of observables can be calculated simultaneously by simply
histogramming the quantities of interest, experimental acceptance cuts can
easily be imposed on the calculation, and
it is also
possible to compute the NLO QCD corrections for exclusive channels,
{\it e.g.}, $p\,p\hskip-7pt\hbox{$^{^{(\!-\!)}}$} \rightarrow V_1 V_2+0$~jet.
Helicity amplitude methods make cross section calculations tractable for
processes involving a large number of tree level diagrams, futhermore,
the leptonic decays of the weak bosons are trivial to implement at the
amplitude level.

The remainder of this paper is organized as follows.  The formalism used in the
calculations is described in Sec.~II, numerical results for the Tevatron and
LHC
center of mass energies are given in Sec.~III, and summary remarks
are given in Sec.~IV.  Technical details of the calculation have been relegated
to an appendix.

\section{FORMALISM}

Next-to-leading-order
calculations of hadronic $ZZ$, $W^- W^+$, and $W^{\pm}Z$ production
have been presented for real weak bosons in the final state.
These results are extended in this section
to include the leptonic decays $W \to \ell \nu$ and $Z \to \ell \bar \ell$
$(\ell = e, \mu)$.
This section begins with a brief review
of the NLO Monte Carlo formalism used in the calculation
and concludes with a discussion of the incorporation of the leptonic decays.

The calculations are done using the narrow width approximation for
the leptonically decaying weak bosons.
This simplifies the calculation greatly for two
reasons.  First of all, it is possible to ignore the contributions from
non-resonance Feynman diagrams without violating gauge invariance.
An example of such a diagram is $q \bar q \to Z \to e \bar e$
followed by $\bar e \to \bar e Z \to \bar e \mu \bar \mu$,
yielding the final state $e\bar e \mu \bar \mu$,
which is the same final state
produced by $q \bar q \to ZZ \to e\bar e \mu \bar \mu$.
Secondly, in the narrow width approximation it is particularly easy to extend
the NLO calculations of real weak boson pairs
to include the leptonic decays of the
$W$- and $Z$-bosons.

\subsection{Monte Carlo Formalism}

The NLO calculations of $V_1 V_2$ production include contributions from the
square of the Born diagrams, the interference
between the Born diagrams and the virtual one-loop diagrams,
and the square of the real emission diagrams.  (The Feynman
diagrams for the case of real weak bosons in the final state
can be found in the original references.)
The calculations have been done using a combination of analytic
and Monte Carlo integration
methods~\cite{NLOMC}. The basic idea is to isolate the
soft and collinear singularities associated with the real emission
subprocesses by partitioning phase space into soft, collinear, and
finite regions.  This is done by introducing theoretical soft and
collinear cutoff parameters, $\delta_s$ and $\delta_c$.  Using
dimensional regularization~\cite{TV},  the soft and collinear
singularities are exposed as poles in  $\epsilon$ (the number of
space-time dimensions is $N = 4 - 2\epsilon$ with $\epsilon$ a small
number).  The infrared singularities from the soft and virtual
contributions are then explicitly canceled while the collinear
singularities are factorized and absorbed into the definition of the
parton distribution functions.
The remaining contributions are finite
and can be  evaluated in four dimensions.  The Monte Carlo program thus
generates $n$-body (for the Born and virtual contributions) and
$(n+1)$-body (for the real emission contributions) final state events.
The $n$- and $(n+1)$-body contributions both depend on the cutoff
parameters $\delta_s$ and $\delta_c$, however, when these contributions
are added together to form a suitably inclusive observable, all
dependence on the cutoff parameters cancels.
The numerical results presented in this paper are insensitive to
variations of the cutoff parameters.

\subsection{Summary of ${\cal O}(\alpha_s)$ $V_1 V_2$ production followed by
leptonic decays}

The formalism for ${\cal O}(\alpha_s)$ hadronic weak boson pair production
followed by leptonic decays of the weak bosons is summarized here.
A detailed discussion about the incorporation of leptonic decays into a NLO
calculation of a real weak boson pair can be found in Ref.~\cite{WGAMMANS},
where the leptonic decay $W \to e \nu$ was
incorporated into a NLO calculation of real
$W\gamma$ production.
Basically, except for the virtual contribution, all the NLO contributions for
real $V_1 V_2$ production have the form
\widetext
\begin{eqnarray}
d\sigma^{\hbox{\scriptsize NLO}}  (q_1 \bar q_2 \to V_1 V_2) =
d\sigma^{\hbox{\scriptsize Born}} (q_1 \bar q_2 \to V_1 V_2) \>
\left[ 1 + C_F \, {\alpha_s \over 2 \pi} (\ \ldots \ ) \right] \>,
\label{EQ:NLOFORM}
\end{eqnarray}
\narrowtext
where $\sigma^{\hbox{\scriptsize Born}}$ is the lowest order Born contribution,
$C_F = 4/3$ is the quark-gluon vertex color factor, and $\alpha_s$ is the
strong
running coupling.
Thus the leptonic decays can be incorporated by simply making the replacement
\widetext
\begin{eqnarray}
d\sigma^{\hbox{\scriptsize Born}} (q_1 \bar q_2 \to V_1 V_2) \longrightarrow
d\sigma^{\hbox{\scriptsize Born}} (q_1 \bar q_2 \to V_1 V_2
\to \ell_1 \bar\ell_1 \ell_2 \bar \ell_2) \>
\label{EQ:NLOREPLACE}
\end{eqnarray}
\narrowtext
in the formulas for NLO real $V_1 V_2$ production.
The leptonic decays are particularly easy to incorporate when the calculation
is
done at the amplitude level; the weak boson polarization
vectors, $\epsilon_\mu (k)$, are simply replaced by the $V\to \ell \bar \ell$
decay currents, $J_\mu (k)$, in the
amplitude.  Details of the amplitude level calculations for the Born and real
emission subprocesses can be found in Ref.~\cite{VVJETTWO}.

The simple replacement described in the previous paragraph does not hold for
the virtual corrections.
Rather than undertake the non-trivial task of
recalculating the virtual corrections for the case of
leptonically decaying weak bosons, we have instead opted to use the
virtual corrections for real on-shell
weak bosons which we subsequently decay ignoring spin correlations.
Neglecting spin correlations slightly modifies the shapes of the angular
distributions of the final state leptons, but the total cross sections are not
altered as long as no angular cuts ({\it e.g.}, rapidity cuts) are imposed on
the final state leptons.  For realistic rapidity cuts, the total cross
sections are changed by typically 10\% when spin correlations are neglected.
Since the virtual corrections are small (they are typically
less than
10\% as large as the corresponding Born cross section)
and the effects of spin correlations are
small, the overall result of ignoring spin correlations in the virtual
corrections is negligible compared to the 20\% -- 30\% uncertainty from the
parton distribution functions and the choice of the scale $Q^2$.
(Note that spin correlations are included everywhere in the calculations
except in the virtual contributions.)

The results for the NLO calculation of
$p\,\bar p \rightarrow V_1 V_2 + X
\rightarrow \ell_1 \bar\ell_1 \ell_2 \bar \ell_2 + X$
can now be summarized.
(The same formalism holds for $p\,p$ collisions with the obvious replacement
$\bar p \to p$.)
The NLO cross section consists of four-
and five-body final state contributions:
\widetext
\begin{eqnarray}
\noalign{\vskip 5pt}
\sigma^{\hbox{\scriptsize NLO}}
 (p\,\bar p \to V_1 V_2 + X \to \ell_1 \bar\ell_1
\ell_2 \bar \ell_2 + X) = &\phantom{+}&
\sigma^{\hbox{\scriptsize NLO}}_{\hbox{\scriptsize 4 body}}
(p\,\bar p \to V_1 V_2 \to \ell_1 \bar\ell_1
\ell_2 \bar \ell_2) \label{EQ:NLOCROSSDECAY} \\
\noalign{\vskip 5pt}
&+& \sigma_{\hbox{\scriptsize 5 body}}
(p\,\bar p \to V_1 V_2 + X \to \ell_1 \bar\ell_1
\ell_2 \bar \ell_2 + X) \>.
\nonumber
\end{eqnarray}
\narrowtext
The four-body contribution is
\widetext
\begin{eqnarray}
\noalign{\vskip 5pt}
& &\sigma^{\hbox {\scriptsize NLO}}_{\hbox{\scriptsize 4 body}}
 (p\,\bar p \to V_1 V_2
\to \ell_1 \bar\ell_1 \ell_2 \bar \ell_2) =
 \sigma^{\hbox{\scriptsize hc}}
 + \sum_{q_1, \bar q_2} \int dv\,dx_1\,dx_2  \label{EQ:TWELVE} \\
\noalign{\vskip 5pt}
 & &\quad \times \biggl[ G_{q_1/p}(x_1,M^2) \,
		         G_{\bar q_2/ \bar p} (x_2,M^2) \,
 {d\hat\sigma^{{\hbox {\scriptsize NLO}}}\over dv}
 (q_1\bar q_2\rightarrow V_1 V_2 \to \ell_1 \bar\ell_1 \ell_2 \bar \ell_2)
 + (x_1 \leftrightarrow x_2) \biggr] \>,
\nonumber
\end{eqnarray}
\narrowtext
where $\sigma^{\hbox{\scriptsize hc}}$
is the contribution from the
hard collinear remnants (see the Appendix for the definition),
the sum is over all contributing quark flavors,
$v$ is related to the center of
mass scattering angle $\theta^*$ by $v = {1\over 2}(1 + \cos\theta^*)$,
$x_1$ and $x_2$ are the parton momentum fractions, $G_{q/p}(x,M^2)$ is a parton
distribution function, $M^2$ is the factorization scale, and
\begin{eqnarray}
\noalign{\vskip 5pt}
{d\hat\sigma^{{\hbox {\scriptsize NLO}}}\over dv}
 (q_1\bar q_2 \rightarrow V_1 V_2 \to \ell_1 \bar\ell_1 \ell_2 \bar \ell_2)
&=&
 {d\hat\sigma^{\hbox{\scriptsize Born}}\over dv}
 (q_1 \bar q_2 \to V_1 V_2 \to \ell_1 \bar\ell_1 \ell_2 \bar \ell_2)
\label{EQ:NEWNLO} \\
\noalign{\vskip 5pt}
&\times& \,
\Biggl[ 1 + C_F {\alpha_s(\mu^2) \over 2 \pi} \biggl\{ 4 \ln(\delta_s)^2
+ 3 \ln\Bigl({\hat s \over M^2}\Bigr)
+ 4 \ln(\delta_s) \ln\Bigl({\hat s \over M^2}\Bigr)
 \nonumber \\
\noalign{\vskip 5pt}
&+& \lambda_{FC} \Bigl( 9 + {2\over 3} \pi^2 + 3 \ln(\delta_s)
- 2 \ln(\delta_s)^2 \Bigr) \biggr\} \Biggl] \nonumber \\
\noalign{\vskip 5pt}
&+& {d\hat\sigma^{\hbox{\scriptsize virt}}\over dv}
(q_1 \bar q_2 \to V_1 V_2) \, B(V_1 \to \ell_1 \bar \ell_1) \,
                              B(V_2 \to \ell_2 \bar \ell_2) \>.
\nonumber
\end{eqnarray}
The hat denotes a parton level cross section,
$\alpha_s(\mu^2)$ is the strong running coupling evaluated at
the renormalization scale $\mu^2$, $\delta_s$ is the soft cutoff
parameter, $\hat s$ is the square of the parton center of mass energy,
and $\lambda_{FC}$
specifies the factorization convention: $\lambda_{FC} = 0$ for the universal
(Modified Minimal Subtraction ${\rm \overline{MS}}$~\cite{MSBAR}) convention
and $\lambda_{FC} = 1$ for the physical (Deep Inelastic Scattering DIS)
convention.  The virtual contribution \cite{CORRECTION},
$d\hat\sigma^{\hbox{\scriptsize virt}} / dv (q_1 \bar q_2 \to V_1 V_2)$,
is multiplied by the
$V_1 \to \ell_1 \bar \ell_1$ and
$V_2 \to \ell_2 \bar \ell_2$ branching ratios.

The five-body contribution is
\widetext
\begin{eqnarray}
\noalign{\vskip 5pt}
\sigma^{\phantom{\hbox{\scriptsize NLL}}}_{\hbox{\scriptsize 5 body}}
(p\,\bar p &\to& V_1 V_2 + X \to \ell_1 \bar\ell_1
\ell_2 \bar \ell_2 + X) =
\sum_{a,b,c} \int d\hat\sigma(ab \to V_1 V_2 c \to
\ell_1 \bar\ell_1 \ell_2 \bar \ell_2 c) \\
\noalign{\vskip 5pt}
&\times& \Bigl[ G_{a/p}(x_1,M^2)\,G_{b/ \bar p} (x_2,M^2)
 + (x_1 \leftrightarrow x_2) \Bigr] dx_1 \,dx_2 \>,
\nonumber
\end{eqnarray}
\narrowtext
where the sum is over all partons contributing to the three subprocesses
$q_1 \bar q_2 \to V_1 V_2 g   \to \ell_1 \bar\ell_1 \ell_2 \bar \ell_2 g$,
$q_1 g        \to V_1 V_2 q_2 \to \ell_1 \bar\ell_1 \ell_2 \bar \ell_2 q_2$,
and
$g\bar q_2\to V_1 V_2\bar q_1\to\ell_1\bar\ell_1\ell_2\bar\ell_2\bar q_1$.
The $2 \to 5$ subprocess is labeled by
$p_1 + p_2 \to p_3 + p_4 + p_5 + p_6 + p_7$ and the
kinematic invariants $s_{ij}$ and $t_{ij}$ are defined by
$s_{ij} = (p_i + p_j)^2$ and $t_{ij} = (p_i - p_j)^2$.
The integration over five-body phase space and $dx_1 \, dx_2$
is done numerically by standard Monte Carlo techniques.  The
kinematic invariants $s_{ij}$ and $t_{ij}$ are first tested for soft
and  collinear singularities.  If an invariant for a subprocess falls
in a soft  or collinear region of phase space, the contribution from
that subprocess is  not included in the cross section.
The squared matrix elements for the Born and real emission
subprocesses were evaluated numerically via helicity amplitude methods as
described in Ref.~\cite{VVJETTWO}.
No attempt has been made to antisymmeterize the amplitudes for the case of
identical fermions in the final state.

\section{PHENOMENOLOGICAL RESULTS}

The phenomenological implications of NLO QCD
corrections to $V_1 V_2$ production at the Tevatron
($p\,\bar p$ collisions at $\sqrt{s} = 1.8$~TeV) and the LHC ($p\,p$
collisions at $\sqrt{s} = 14$~TeV) will now be discussed.
This section begins with a brief description of
the input parameters and acceptance cuts.

\subsection{Input Parameters}

The numerical results presented in this
section were obtained using the two-loop expression for
$\alpha_s$. The QCD scale $\Lambda_{\hbox{\scriptsize QCD}}$
is specified for four
flavors of quarks by the choice of the parton distribution functions and
is adjusted whenever a heavy quark threshold is crossed so that
$\alpha_s$ is a continuous function of $Q^2$. The heavy quark masses
were taken to be $m_b=5$~GeV and $m_t=150$~GeV.
The SM parameters used in the numerical simulations are $M_Z = 91.173$~GeV,
$M_W = 80.22$~GeV, $\alpha (M_W) =1/128$, and $\sin^2
\theta_{\hbox{\scriptsize w}} = 1 - (M_W^{}/M_Z^{})^2$. These values are
consistent with recent measurements at LEP, the CERN $p\,\bar p$
collider, and the Tevatron \cite{LEP,MT,MW}. The soft and collinear
cutoff parameters are fixed to $\delta_s = 10^{-2}$ and $\delta_c =
10^{-3}$ unless stated otherwise. The parton
subprocesses have been summed over $u,d,s$, and $c$ quarks and the
Cabibbo mixing angle has been chosen such that $\cos^2 \theta_C =
0.95$. The leptonic branching ratios are
$B(W \to e \nu) = 0.107$ and $B(Z \to e \bar e) = 0.034$
and the total widths of the $W$- and $Z$-bosons are
$\Gamma_W = 2.12$~GeV and $\Gamma_Z = 2.487$~GeV.
A single scale
$Q^2=M^2_{V_1 V_2}$,  where $M_{V_1 V_2}$ is the invariant mass of the
$V_1 V_2$ pair, has been used for the renormalization scale $\mu^2$
and the factorization scale $M^2$.

In order to get consistent NLO results it is necessary to use parton
distribution functions which have been fit to next-to-leading order.
The numerical results were obtained using the
the Martin-Roberts-Stirling (MRS)~\cite{MRSPRIME} set S0$^\prime$
distributions
with $\Lambda_4 = 215$~MeV.
The MRS distributions are defined in the universal (${\rm \overline{MS}}$)
scheme and thus the factorization defining parameter $\lambda_{FC}$ in
Eqs.~(\ref{EQ:NEWNLO}) and (\ref{EQ:AP})
should be $\lambda_{FC} = 0$. For convenience,
the MRS set S0$^\prime$
distributions have also been used for the leading order (LO) calculations.

\subsection{Cuts}

The cuts imposed in the numerical simulations are motivated by
the finite acceptance and resolution of the detector. The finite
acceptance of the detector is simulated by cuts on the four-vectors of the
final state particles.  These cuts include requirements on the
transverse momentum, $p_T^{}$,
and pseudorapidity, $\eta = \ln\cot(\theta / 2)$,
of the charged leptons and on the missing
transverse momentum, $p\llap/_T^{}$, associated with the neutrino(s).
Charged leptons are also required to be separated in
azimuthal angle-pseudorapidity space,
$\Delta R = [(\Delta \phi)^2 +  (\Delta \eta)^2 ]^{1/2}$,
where $\theta$ and $\phi$ are the polar and azimuthal angles relative to the
beam; this cut is impelled by the finite granularity of the detector.
The complete set of cuts can be summarized as follows.
\begin{quasitable}
\begin{tabular}{cc}
Tevatron & LHC\\
\tableline
$p_{T}^{}(\ell)         >  20$~GeV & $p_{T}^{}(\ell)         > 25$~GeV\\
$p\llap/_T^{}           >  20$~GeV & $p\llap/_T^{}           > 50$~GeV\\
$|\eta(\ell)|           < 2.5$     & $|\eta(\ell)|           < 3.0$\\
$\Delta R(\ell,\ell)    < 0.4$     & $\Delta R(\ell,\ell)    < 0.4$\\
\end{tabular}
\end{quasitable}

\subsection{NLO Cross Sections}

The dependence of the total cross section on the soft and collinear
cutoff parameters is illustrated in Fig.~\ref{FIG:DELTASC} which shows
the total NLO cross section for
$p\,p \to ZZ + X \to e^- e^+ \mu^- \mu^+ + X$
plotted versus $\delta_s$ and $\delta_c$, for $\sqrt{s} =
14$~TeV and the cuts described in Sec.~IIIB. The $n$- and $n+1$-body
contributions are also plotted for illustration ($n=4$ for this process).
The figure shows that the $4$- and $5$-body
contributions, which separately have no physical meaning, vary strongly with
$\delta_s$ and $\delta_c$, however, the total cross section, which is the sum
of the $4$- and $5$-body contributions, is independent of $\delta_s$
and $\delta_c$ over a wide range of these parameters.

The total LO and NLO cross sections for weak boson pair production, with the
cuts specified in Sec.~IIIB, are given in Table~1 for center of mass energies
corresponding to the present Tevatron ($\sqrt{s} = 1.8$~TeV), an upgraded
Tevatron ($\sqrt{s} = 3.5$~TeV), and the proposed LHC ($\sqrt{s} = 14$~TeV).
The cross sections in Table~1 have been summed over $\ell = e, \mu$ and
both charges of the $W$ in the $W^{\pm}Z$ process.

For definiteness and convenience, differential cross
sections will be given for the processes
$p\,p\hskip-7pt\hbox{$^{^{(\!-\!)}}$} \to Z Z + X
\to e^- e^+  \mu^- \mu^+ + X$,
$p\,p\hskip-7pt\hbox{$^{^{(\!-\!)}}$} \to W^- W^+ + X
\to e^- \bar \nu_e \nu_e e^+ + X$, and
$p\,p\hskip-7pt\hbox{$^{^{(\!-\!)}}$} \to W^+ Z + X
\to \nu_e e^+ \mu^- \mu^+ + X$.
In practice, the final state leptons would be summed over $\ell = e, \mu$
and both charges of the $W$ would be summed in the $WZ$ process.
Thus the distributions given here would be scaled up by the
appropriate power of 2.
The differential cross sections include the cuts described in Sec.~IIIB.
The figures are arranged in two parts, with parts a) and b) being
the results for the Tevatron ($\sqrt{s} = 1.8$~TeV) and
LHC ($\sqrt{s} = 14$~TeV) energies, respectively.
Both NLO and LO results are shown.  In some distributions, the NLO 0-jet
exclusive and the LO 1-jet exclusive cross sections are also given.  For these
exclusive cross sections it is necessary to define a jet.  A jet will be
defined as a final state quark or gluon with
\begin{eqnarray}
p_T^{}(j)>10~{\rm GeV}\hskip 1.cm {\rm and} \hskip 1.cm |\eta(j)|<2.5
\label{EQ:TEVJET}
\end{eqnarray}
at the Tevatron, and
\begin{eqnarray}
p_T^{}(j)>50~{\rm GeV}\hskip 1.cm {\rm and} \hskip 1.cm |\eta(j)|<3
\label{EQ:SSCJET}
\end{eqnarray}
at the LHC. The sum of the NLO 0-jet and the LO 1-jet exclusive cross
sections is equal to the inclusive NLO cross section.

\subsection{$ZZ$ production}

The first process to be considered is
$p\,p\hskip-7pt\hbox{$^{^{(\!-\!)}}$} \to Z Z + X
\to e^- e^+  \mu^- \mu^+ + X$.
The invariant mass distribution of the four leptons
is displayed in Fig.~\ref{FIG:ZZ:M}; NLO and LO cross sections
are shown.
The NLO corrections are nearly uniform in the
invariant mass at the Tevatron energy and increase only slightly with the
invariant mass at the LHC energy.

Figure~\ref{FIG:ZZ:PTL} shows the inclusive differential cross section for the
lepton transverse momentum (all four leptons have been histogrammed,
each with the full event weight).
At the Tevatron energy, the NLO corrections increase slowly with
$p_T^{}(\ell)$, whereas at the LHC energy, the corrections increase more
rapidly with $p_T^{}(\ell)$.
The 0-jet and 1-jet exclusive cross
sections are also shown.  At the Tevatron energy, the 0-jet exclusive cross
section is slightly larger than the LO cross section, while the 1-jet exclusive
cross section is much smaller than the LO cross section.  At the LHC energy,
the 1-jet exclusive cross section is much smaller than the LO cross section at
small values of $p_T^{}(\ell)$, but becomes comparable to the LO cross section
at large values of $p_T^{}(\ell)$.
The decomposition of the NLO cross section into 0-jet and 1-jet components
shows that the large NLO corrections at high $p_T^{}(\ell)$ are due to
contributions from 1-jet real emission subprocesses.
The 1-jet exclusive cross section becomes a larger fraction of the total  NLO
cross section at higher energies because the contributions from $qg$ initial
state processes grow with the center of mass energy due to the
increasing gluon luminosity.

The 0-jet and 1-jet exclusive cross sections are of course arbitrary since they
depend on the jet definition.  For example, increasing the $p_T^{}$ threshold
for a jet will suppress the 1-jet and enhance the 0-jet exclusive cross
sections (the 0-jet and 1-jet exclusive cross sections must sum to the NLO
cross section).  Nevertheless, Figure~\ref{FIG:ZZ:PTL} illustrates that for
reasonable jet definitions, the 1-jet contribution to the inclusive NLO cross
section is small at the Tevatron energy, but becomes significant at the LHC
energy, especially at high $p_T^{}(\ell)$.

The inclusive differential cross section for the lepton pseudorapidity,
$\eta = \ln \cot(\theta/2)$ where $\theta$ is the polar angle of the lepton
with respect to the proton direction in the laboratory frame, is given in
Fig.~\ref{FIG:ZZ:YL}.  The NLO corrections are largest in the central rapidity
region.  The 0-jet and 1-jet exclusive cross sections are also shown.
Notice that in the $ZZ$ process, the 1-jet exclusive cross section is
small compared to the NLO cross section.
For the $W^- W^+$ and $WZ$ processes, the 1-jet exclusive cross section becomes
a larger fraction of the total NLO cross section.

The angular distributions of the leptonic decay products contain information on
the helicities of the vector bosons.  These distributions are simplest in the
rest frame of the individual vector bosons.
For the decay of a polarized $Z$-boson, $Z \to e^- e^+$, the angular
distributions of the $e^-$ in the $Z$-boson rest frame are
\begin{eqnarray}
{d\Gamma \over d\cos\theta} (\lambda_{Z}^{} = 0) &=&
{G_F M_Z^3 \over 2 \pi \sqrt{2} } \ (g_V^2 + g_A^2) \> \sin^2 \theta \>,
\label{EQ:ZZERO} \\
\noalign{\vskip 5pt}
{d\Gamma \over d\cos\theta} (\lambda_{Z}^{} = \pm 1) &=&
{G_F M_Z^3 \over 2 \pi \sqrt{2} } \
\Bigl[ (g_V^2 + g_A^2) \> {1\over 2} (1 + \cos^2 \theta)
\pm 2 g_V^{} g_A^{} \cos\theta \Bigr] \>,
\label{EQ:ZPM}
\end{eqnarray}
where $\theta$ is the angle of the $e^-$ with respect to the longitudinal axis
and $\lambda_{Z}^{}$ denotes the polarization of the $Z$-boson;
$\lambda_{Z}^{} = 0$ and $\lambda_{\rm z} = \pm 1$ denote the
longitudinal and transverse polarizations, respectively.
Here $G_F$ is the Fermi coupling constant and $g_V^{}$ ($g_A^{}$) is the
vector (axial vector) coupling of the $Z$-boson to fermions.
Since the $Z$-boson coupling to charged leptons is almost purely axial vector,
transversely polarized $Z$-bosons produce a ${1\over 2}(1 + \cos^2 \theta)$
distribution for the $e^-$, while longitudinally polarized $Z$-bosons
yield a $\sin^2 \theta$ distribution.
Figure~\ref{FIG:ZZ:COSE} shows the polar angle distribution of the $e^-$
in the parent $Z$-boson rest frame, measured
with respect to the parent $Z$-boson direction in the $ZZ$ rest frame,
{\it i.e.}, $\cos\theta_{e^-} = \hat p_{e^-} \cdot \hat p_{Z}^{}$ where
$\hat p_{e^-}$ is the unit-normalized three-momentum
of the $e^-$ in the parent $Z$-boson
rest frame and $\hat p_{Z}^{}$ is the unit-normalized three-momentum
of the parent $Z$-boson in the $ZZ$ rest frame.
The shapes of the distributions in Fig.~\ref{FIG:ZZ:COSE} indicate that the
transverse polarizations are dominating the cross section.  This is to be
expected since the $q \bar q$ annihilation process produces $Z$-boson pairs
that are primarily transversely polarized, especially at
large parton-center-of-mass energies.  The sharp drops in the distributions
near $\cos \theta_{e^-} = \pm 1$ are due to the kinematic cuts.

Figure~\ref{FIG:ZZ:COSCHI} shows the angular correlation between the decay
planes formed by the leptons.  The angle $\chi$ between the decay planes
is defined by
\begin{equation}
\cos \chi = { (\vec p_{e^-}   \times \vec p_{e^+}   ) \cdot
              (\vec p_{\mu^-} \times \vec p_{\mu^+} ) \over
	      |\vec p_{e^-}   \times \vec p_{e^+}   | \ \
              |\vec p_{\mu^-} \times \vec p_{\mu^+} | }\>,
\label{EQ:CHI}
\end{equation}
where the momentum vectors are defined in the $ZZ$ rest frame.  The NLO
and LO curves have the same shape at both energies.
The shapes of these curves are dominated by the effects of the kinematic cuts;
without cuts the curves are essentially flat \cite{KANE}.

\subsection{$W^-W^+$ production}

Attention now turns to the process
$p\,p\hskip-7pt\hbox{$^{^{(\!-\!)}}$} \to W^- W^+ + X
\to e^- \bar \nu_e \nu_e e^+ + X$.  Since there are two invisible neutrinos
in the final state,
much of the final state kinematic information is lost and
it is impossible to reconstruct the $W^-W^+$ invariant mass.
The best one can do is form the transverse
cluster mass \cite{CTMASS} defined by
\widetext
\begin{eqnarray}
M_T^2(c,p\llap/_T^{}) = \biggl[ \sqrt{p^2_{cT} + m_c^2}
+ \left| p\llap/_T^{} \right| \biggr]^2
- \biggl[\vec{p}_{cT}^{} + \vec{p}\llap/_T^{} \biggr]^2 \>,
\label{EQ:CTM}
\end{eqnarray}
\narrowtext
where $c$ is either a single particle or a cluster of several particles.
For the $W^- W^+$ process the cluster is $c = e^- + e^+$.
The transverse cluster mass distribution is displayed in Fig.~\ref{FIG:WW:CTM}.
The NLO corrections are nearly uniform in
$M_T^{}$ at both the Tevatron and LHC energies.
The invariant mass of the charged leptons, $M(e^- e^+)$, is also unchanged
in shape by the NLO corrections.

Figure~\ref{FIG:WW:PTL} shows the inclusive differential cross
section for the charged lepton transverse momentum
(both charged leptons have been histogrammed,
each with the full event weight).
The NLO corrections increase slowly with $p_T^{}(\ell)$
at the Tevatron energy, but at the LHC energy,
they increase very rapidly with
$p_T^{}(\ell)$.  Similar behavior is observed in the $p_T^{}$
spectra of the $W$-bosons \cite{WW,WWFRIX}.
The 0-jet and 1-jet exclusive cross sections are also shown.  At the Tevatron
energy, the 0-jet cross section is always larger than the 1-jet cross section.
At the LHC energy, on the other hand, the 0-jet cross section dominates at
small $p_T^{}(\ell)$, while the 1-jet cross section dominates at high
$p_T^{}(\ell)$.  This behavior is similar to that observed in the $ZZ$ process,
except now the 1-jet component is a larger fraction of the total NLO cross
section.

The missing transverse momentum distribution is presented in
Fig.~\ref{FIG:WW:PTMISS}. The $p\llap/_T^{}$ distribution begins to fall
rapidly when $p\llap/_T^{} \approx M_W^{}$.  The NLO corrections increase with
$p\llap/_T^{}$ and become very large, especially at the LHC energy.
The 0-jet and 1-jet cross
sections are also shown.  The 0-jet cross section dominates for
$p\llap/_T^{} \lesssim M_W^{}$ while the 1-jet cross section dominates for
$p\llap/_T^{} > M_W^{}$.  The large NLO corrections at high
$p\llap/_T^{}$ are due to 1-jet events.
At LO the $W$-bosons are back-to-back in the transverse plane, thus when
high $p_T^{}$ $W$-bosons decay, the decay product neutrinos will also be nearly
back-to-back in the transverse plane.  The missing transverse momentum, which
is the vector sum of the neutrino transverse momenta, thus tends to be small
due to the acollinear cancellation of the neutrino momenta
and the $p\llap/_T^{}$ distribution falls
rapidly.  On the other hand, when a $W^- W^+$ event contains a high $p_T^{}$
jet, the transverse angle between the decay product neutrinos can easily be
acute, thus yielding a much larger value of $p\llap/_T^{}$.

Figure~\ref{FIG:WW:YL} shows the inclusive pseudorapidity distribution of the
charged leptons.  The NLO corrections are once again largest in the central
pseudorapidity region.  The 0-jet and 1-jet exclusive cross sections are also
shown.  The interesting feature to note is that compared to
the corresponding distribution for the $ZZ$ process
(Fig.~\ref{FIG:ZZ:YL}), the 1-jet cross section for the $W^-W^+$ process
is a larger fraction of the total NLO cross section.  This trend will
continue for the $WZ$ process.

\subsection{$WZ$ production}

The final process to be considered is
$p\,p\hskip-7pt\hbox{$^{^{(\!-\!)}}$} \to W^+ Z + X
\to \nu_e e^+ \mu^- \mu^+ + X$.
This process is of special interest because it is sensitive to the $WWZ$
vertex.  The discussion here will be limited to the case of standard model
couplings at the $WWZ$ vertex.
A study of this process with NLO corrections and anomalous
couplings at the $WWZ$ vertex can be found in Ref.~\cite{WZNONSM}.

The transverse cluster mass [see Eq.~(\ref{EQ:CTM})],
where the cluster is $c = e^+ + \mu^- + \mu^+$, is
shown in Fig.~\ref{FIG:WZ:CTM}.  The NLO corrections are nearly uniform in
$M_T^{}$ at both the Tevatron and LHC energies.
Since there is only one neutrino in the final state, it is possible to
reconstruct the $WZ$ invariant mass by requiring the invariant mass of the
electron plus neutrino system to be equal to the $W$-boson mass.
This constraint gives a quadratic solution for the
longitudinal momentum of the
neutrino and thus there is a two-fold ambiguity in the reconstructed $WZ$
invariant mass. The reconstructed
$WZ$ invariant mass, formed by histogramming both reconstructed invariant
masses, each with half the event weight,
is qualitatively similar to the cluster transverse mass.
In particular, the shape of the reconstructed $WZ$ invariant mass is unchanged
by the NLO corrections.

Figure~\ref{FIG:WZ:PTL} shows the inclusive transverse momentum distribution of
the charged leptons
(all three charged leptons have been histogrammed,
each with the full event weight).
The NLO
corrections once again increase with $p_T^{}(\ell)$ and are especially large at
the LHC energy.  The 0-jet exclusive cross section is the dominant component of
the inclusive NLO cross section at the Tevatron energy, whereas at the LHC,
the 0-jet
cross section dominates at small $p_T^{}(\ell)$  while the 1-jet cross
section dominates at high $p_T^{}(\ell)$.
The $p\llap/_T^{}$ distribution, which is shown in Fig.~\ref{FIG:WZ:PTMISS},
exhibits the same qualitative features as the
$p_T^{}(\ell)$ distribution.

The inclusive pseudorapidity distribution of the charged leptons is displayed
in Fig.~\ref{FIG:WZ:YL}.  Note that the 1-jet cross section is now an even
larger fraction of the total NLO cross section than it was in either the $ZZ$
or $W^-W^+$ process (see Figs.~\ref{FIG:ZZ:YL} and ~\ref{FIG:WW:YL}).

Figure~\ref{FIG:WZ:COSE} shows the polar angle distribution of the $e^+$
in the $W$-boson rest frame, measured
with respect to the $W$-boson direction in the
$W^+ Z$ rest frame,
{\it i.e.}, $\cos \theta_{e^+} = \hat p_{e^+} \cdot \hat p_{W}^{}$ where
$\hat p_{e^+}$ is the unit-normalized
three-momentum of the $e^+$ in the $W$-boson
rest frame and $\hat p_{W}^{}$ is the unit-normalized three-momentum of
the $W$-boson in the $WZ$ rest frame.
The $\cos \theta_{e+}$ distribution, and the analogous $\cos \theta_{\mu^-}$
distribution discussed in the next paragraph, both contain two-fold ambiguities
corresponding to the two solutions
for the longitudinal momentum of the neutrino.
Both solutions have been histogrammed, each with
half the event weight.
For the decay of a polarized $W$-boson, $W^+ \to e^+ \nu_e$, the angular
distributions of the $e^+$ in the $W$-boson rest frame are
\begin{eqnarray}
{d\Gamma \over d\cos\theta} (\lambda_{W}^{} = 0) &=&
{G_F M_W^3 \over 8 \pi \sqrt{2} } \ \sin^2 \theta \>, \\
\noalign{\vskip 5pt}
{d\Gamma \over d\cos\theta} (\lambda_{W}^{} = \pm 1) &=&
{G_F M_W^3 \over 8 \pi \sqrt{2} } \ {1\over 2} (1 \pm \cos\theta)^2 \>,
\end{eqnarray}
where $\theta$ is the angle of the $e^+$ with respect to the longitudinal axis
and $\lambda_{W}^{} = 0$ ($\pm 1$) denotes the
longitudinal (transverse) polarization(s) of the $W^+$-boson.
The $q \bar q$ annihilation process produces $WZ$ pairs that are primarily
transversely polarized, futhermore, for $q_1 \bar q_2 \to W^+ Z$, the helicity
combination $(\lambda_{W}^{} = -1, \lambda_{Z}^{} = 1)$ gives the dominant
contribution \cite{WZZERO}.
This explains the ${1\over 2}(1-\cos\theta)^2$ shape of the
$\cos\theta_{e^+}$ distribution in Fig.~\ref{FIG:WZ:COSE}.
The distributions fall near $\cos\theta_{e^+} = \pm 1$ due to the kinematic
cuts.

Figure~\ref{FIG:WZ:COSMU} shows the polar angle distribution of the $\mu^-$
in the $Z$-boson rest frame, measured with respect to the $Z$-boson
direction in the $W^+ Z$ rest frame,
{\it i.e.},  $\cos\theta_{\mu^-} = \hat p_{\mu^-} \cdot \hat p_{Z}^{}$ where
$\hat p_{\mu^-}$ is the unit-normalized three-momentum
of the $\mu^-$ in the $Z$-boson rest frame and $\hat p_{Z}^{}$ is the
unit-normalized three-momentum
of the $Z$-boson in the $WZ$ rest frame.
Vestiges of the
${1\over 2}(1+\cos^2\theta)$ distribution characteristic of transversely
polarized $Z$-bosons can be seen in the figure.
The asymmetry in the distributions
comes from the $\cos\theta$ term in
Eq.~(\ref{EQ:ZPM}) and has a negative slope because
the dominant contribution comes from the helicity
combination $(\lambda_{W}^{} = -1, \lambda_{Z}^{} = 1)$.

\subsection{Discussion}

Comparing the three processes, one sees that the NLO corrections increase for
the processes in the order $ZZ$, $W^- W^+$, $WZ$.  The 1-jet exclusive
cross section also becomes a larger fraction of the total NLO cross section in
this same order.  Both of these features are due to the $qg \to V_1 V_2 q$ real
emission subprocesses.

A decomposition of the NLO corrections into components from order $\alpha_s$
$q\bar q$ and $qg$ initial states shows that the $q\bar q$ components are of
similar size in all three processes (the $q\bar q$ component is approximately
10\% the size of the corresponding Born cross section), whereas the order
$\alpha_s$ $qg$ component increases for the processes in the order
$ZZ$, $W^- W^+$, $WZ$ \cite{WZ,FRIX,WWFRIX}.
A similar decomposition of the 1-jet exclusive cross
section exhibits the same behavior.

The NLO corrections and the 1-jet fraction are largest for the $WZ$ process
because the Born cross section for this process is suppressed due to
destructive interference between the $S$-, $T$-, and $U$-channel Feynman
diagrams.  This destructive interference produces an approximate amplitude
zero \cite{WZZERO};
the dominant helicity amplitudes $(\lambda_W = \pm 1, \lambda_Z = \mp 1)$
have an exact zero, whereas the other helicity amplitudes remain
finite but small.   The situation is analogous to the $q\bar q \to W\gamma$
process which, because of the massless photon, has an exact zero in
all the helicity amplitudes \cite{RAZ}.
The NLO QCD corrections to the process
$q \bar q \to W\gamma$ are very
large at high center of mass energies as a result of
the destructive interference and the large gluon luminosity \cite{WGAMMA}.
The approximate amplitude zero in the $q\bar q \to WZ$ process produces dips in
the distributions of $\cos\theta^*$ \cite{WZZERO}, $y_Z^*$,
and $y_Z^{} - y_W^{}$ \cite{FRIX}, where
$\theta^*$ is the center of mass scattering angle, $y_Z^*$ is the $Z$-boson
rapidity in the $WZ$ center of mass frame, and $y_Z^{}$ ($y_W^{}$) is the
$Z$-boson ($W$-boson) rapidity in the laboratory frame.  The approximate
amplitude zero also suppresses the $WZ$ Born cross section at high $p_T^{}(Z)$.
The $q g \to WZq$ subprocess, on the other hand, is not suppressed by
destructive interference.  As a result, the NLO corrections and the
1-jet fraction
are larger for the $WZ$ process than for either the $ZZ$ or $W^-W^+$ process.

In all three processes, the NLO corrections are largest at high $p_T^{}$
because the $qg \to V_1 V_2 q$ subprocesses are enhanced in diagrams where a
weak boson and a quark are produced at high $p_T^{}$, with the quark
radiating the other weak boson, {\it i.e.}, diagrams in which
$q g \to V_1 q$
followed by $q \to q V_2$.  These subprocesses are enhanced by a factor
$\log^2 (p_T^{}(V_1) / M_{V_2}^{} )$ \cite{FRIX,WWFRIX},
which arises from the kinematic region
where $V_2$ is nearly collinear to the quark.

The $W^-W^+$ process has a larger NLO correction and a larger 1-jet fraction
than the $ZZ$ process.  These features are again due to differences in the
relative importance of the $qg$ initial state processes.
The $qg$ initial state processes are of more relative importance for the
$W^-W^+$ process than for the $ZZ$ process, however, it is not clear why
this is so.   The mass difference between the $W$- and $Z$-boson is not enough
to explain the difference.

Another notable feature of the NLO corrections is that they increase with the
center of mass energy.  This behavior is also due to the $qg$ initial state
processes.  The contributions from these processes increases
with the center of mass energy due to the increasing
the gluon luminosity.

\section{SUMMARY}

The QCD radiative corrections to hadronic $ZZ$, $W^-W^+$,
and $W^{\pm}Z$ production
have been calculated to order $\alpha_s$ with leptonic decays of the weak
bosons included.  The inclusion of the leptonic decays makes the calculations
more realistic since it is the leptonic decay products that are observed in an
experiment.  Distributions of the final state decay products have been given
for both inclusive and exclusive channels for the Tevatron and LHC center of
mass energies.  The calculations include typical acceptance cuts on the final
state leptons.

The calculations were done by using the Monte Carlo method for NLO calculations
in combination with helicity amplitude methods.  With the Monte Carlo method it
is easy to impose experimentally motivated acceptance cuts on the final state
leptons, also, it is possible to calculate the order $\alpha_s$ QCD corrections
for exclusive channels, {\it e.g.},
$p\, p\hskip-7pt\hbox{$^{^{(\!-\!)}}$} \rightarrow V_1 V_2 + 0$~jet.
The narrow width approximation has been used for the decaying weak bosons.
This simplifies the calculation greatly since it is possible to ignore
contributions from non-resonant Feynman diagrams without violating gauge
invariance.   Futhermore, in the narrow width approximation it is
particularly easy to extend previous NLO calculations of real weak boson pairs
to include the leptonic decays of the weak bosons.  Spin correlations are
included everywhere in the calculation except in the virtual contributions
where they can be safely neglected.

The QCD radiative corrections enhance the transverse momenta spectra of the
leptons at high $p_T^{}$, especially at the LHC energy.  These enhancements
are due to the opening of the $qg$ subprocesses at order $\alpha_s$.  The
contribution from these processes increases with the center of mass energy
due to the increasing gluon density in the proton.
Invariant mass and angular distributions are scaled up in magnitude by the
QCD radiative corrections, but undergo little change in shape.

%
\acknowledgements

The author would like to thank U.~Baur, J.~F.~Gunion, T.~Han,
and D.~Zeppenfeld for useful discussions.
This work has been supported in part by Department of Energy grant
\#DE-FG03-91ER40674 and by Texas National Research Laboratory grant
\#RGFY93-330.

%
%

\appendix
\section*{HARD COLLINEAR CORRECTIONS}

The real emission subprocesses
$q_1(p_1) + \bar q_2(p_2) \to V_1 V_2 g \to \ell_1(p_3)
+ \bar\ell_1(p_4) + \ell_2(p_5) + \bar \ell_2(p_6) + g(p_7)$
(and the cross subprocesses) have hard collinear
singularities when $t_{17} \rightarrow 0$ or $t_{27} \rightarrow 0$
$[t_{ij} = (p_i - p_j)^2]$.
These singularities must be factorized and absorbed into the initial
state parton distribution functions. After the factorization is
performed, the contribution from the remnants
of the hard collinear singularities has the form
\widetext
\begin{eqnarray}
\sigma^{\hbox{\scriptsize hc}} &=& \sum_{q_1,\bar q_2} \int
 {\alpha_s \over 2 \pi} \,
 {d\hat\sigma^{\hbox{\scriptsize Born}} \over dv}
 (q_1 \bar q_2 \rightarrow V_1 V_2 \to \ell_1 \bar\ell_1 \ell_2 \bar \ell_2) \,
dv\,dx_1\,dx_2 \label{EQ:HARDCOL} \\
\noalign{\vskip 5pt}
& &\quad \times \Biggl[ \phantom{+}
G_{q_1/p}(x_1,M^2) \int\limits_{x_2}^{1 - \delta_s} \, {dz \over z} \,
G_{\bar q_2/ \bar p} \Bigl({x_2\over z},
M^2\Bigr) \, \tilde P_{qq}(z)  \nonumber \\
\noalign{\vskip 5pt}
& &\quad \phantom{\times \Biggl[}
+ G_{q_1/p}(x_1,M^2) \int\limits_{x_2}^{1} \, {dz \over z} \,
  G_{g/\bar p} \Bigl({x_2\over z},M^2\Bigr) \, \tilde P_{qg}(z) \nonumber \\
\noalign{\vskip 5pt}
& &\quad \phantom{\times \Biggl[ }
 + G_{\bar q_2/ \bar p} (x_2,M^2) \int\limits_
{x_1}^{1 - \delta_s} \, {dz \over z} \,
   G_{q_1/p} \Bigl({x_1\over z},M^2\Bigr) \, \tilde P_{qq}(z)
 \nonumber \\
\noalign{\vskip 5pt}
& &\quad \phantom{\times \Biggl[ }
 + G_{\bar q_2/ \bar p} (x_2,M^2) \int\limits_
{x_1}^{1} \,  {dz \over z} \,
  G_{g/p} \Bigl( {x_1\over z},M^2\Bigr) \, \tilde P_{qg}(z) \Biggr] \>,
\nonumber
\end{eqnarray}
\narrowtext
with
\begin{eqnarray}
\tilde P_{ij}(z) \equiv P_{ij}(z) \, \ln \biggl( {1-z \over z} \,
 \delta_c \, {s_{12} \over M^2} \biggr)
 - P_{ij}^{\prime}(z) - \lambda_{FC} \, F_{ij}(z) \>.
\label{EQ:AP}
\end{eqnarray}
The Altarelli-Parisi splitting functions in $N = 4 - 2 \epsilon$
dimensions for $0 < z < 1$ are
\begin{eqnarray}
P_{qq} (z,\epsilon) &=& C_F \, \left[ {1 + z^2 \over 1 - z} -
 \epsilon (1 - z) \right] \>, \\
\noalign{\vskip 10pt}
P_{qg} (z,\epsilon) &=& {1 \over 2(1 - \epsilon)} \,
 \Biggl[z^2 + (1-z)^2 - \epsilon \Biggr] \>,
\end{eqnarray}
and can be written as
\begin{eqnarray}
P_{ij} (z,\epsilon) = P_{ij} (z) + \epsilon P_{ij}^{\prime} (z) \>,
\end{eqnarray}
which defines the $P_{ij}^{\prime}$ functions. The functions $F_{qq}$
and $F_{qg}$ depend on the choice of factorization  convention and the
parameter $\lambda_{FC}$ specifies the factorization convention;
$\lambda_{FC} = 0$ for the universal
(Modified Minimal Subtraction ${\rm \overline{MS}}$~\cite{MSBAR})
convention and $\lambda_{FC} = 1$ for the physical (Deep Inelastic
Scattering DIS) convention. For the physical convention the
factorization functions are
\begin{eqnarray}
F_{qq} (z) &=& C_F \, \left[ {1+z^2 \over 1-z} \>
\ln \left({1-z \over z} \right)
- {3 \over 2} \, {1 \over 1-z} + 2 z + 3 \right] , \\
\noalign{\vskip 10pt}
F_{qg} (z) &=& {1\over 2} \left[ \left\{ z^2 + (1-z)^2 \right\}
\ln \left( {1-z \over z} \right) + 8z(1-z) -1 \right] .
\label{EQ:ORIGFQG}
\end{eqnarray}
The transformation between the ${\rm \overline{MS}}$
and DIS schemes is discussed in Ref.~\cite{OWENSTUNG}.
The parameter $M^2$ is the factorization scale which must be specified
in the process of  factorizing the collinear singularity. Basically,
it determines how much of the collinear term is absorbed into the
various parton distribution functions.

\newpage
%
%

%
\newpage
%
\widetext
\begin{table}
\caption{Total cross sections for
$p\,p\hskip-7pt\protect{\hbox{$^{^{(\!-\!)}}$}}
\to Z Z + X \to \ell_1 \bar \ell_1 \ell_2 \bar \ell_2 + X$,
$p\,p\hskip-7pt\protect{\hbox{$^{^{(\!-\!)}}$}}
\to W^- W^+ + X \to \ell_1 \bar \nu_1 \nu_2 \bar \ell_2 + X$,
and
$p\,p\hskip-7pt\protect{\hbox{$^{^{(\!-\!)}}$}}
\to W Z + X \to \ell_1 \nu_1 \ell_2 \bar \ell_2 + X$
for center of mass energies corresponding to the present Tevatron,
an upgraded Tevatron, and the proposed LHC.  The cross sections have been
summed over $\ell = e, \mu$ and both charges of the $W$ in the $WZ$ process.
The cuts listed in Sec.~IIIB have been imposed.}
\label{TABLE1}
\begin{tabular}{cclccc}
 \multicolumn{1}{c}{$\protect{\sqrt{s}}$ (TeV)}
&\multicolumn{1}{c}{$p\,p\hskip-7pt\protect{\hbox{$^{^{(\!-\!)}}$}}$}
&\multicolumn{1}{c}{ }
&\multicolumn{1}{c}{$\sigma(ZZ)$ (fb)}
&\multicolumn{1}{c}{$\sigma(W^-W^+)$ (fb)}
&\multicolumn{1}{c}{$\sigma(W^{\pm}Z)$ (fb)}\\
\tableline
1.8 & $p\, \bar p$ & LO  & 4.8 & 190. & 17.\\
    &              & NLO & 6.2 & 260. & 22.\\
\tableline
3.5 & $p\, \bar p$ & LO  & 12. & 440. & 44.\\
    &              & NLO & 14. & 590. & 59.\\
\tableline
14. & $p\, p$      & LO  & 36. & 570. & 43.\\
    &              & NLO & 43. & 960. & 77.\\
\end{tabular}
\end{table}
\newpage
%
%
\begin{figure}
\caption{The dependance of the
total NLO cross section for $p\,p \rightarrow ZZ + X \rightarrow
e^- e^+ \mu^- \mu^+ + X$ at $\protect{\sqrt{s}} = 14$~TeV on the
soft and collinear cutoff parameters.
In part a) the total NLO cross section is plotted
versus the soft cutoff parameter $\delta_s$ for a fixed value of
$\delta_c = 5 \times 10^{-4}$.
In part b) the total NLO cross section is plotted
versus the collinear cutoff parameter
$\delta_c$ for a fixed value of $\delta_s = 10^{-2}$.
The 4- and 5-body contributions are also shown.
The cuts listed in Sec.~IIIB have been imposed.}
\label{FIG:DELTASC}
\end{figure}

\begin{figure}
\caption{The invariant mass of the four leptons in the process
$p\,p\hskip-7pt\hbox{$^{^{(\!-\!)}}$} \to Z Z + X
\to e^- e^+  \mu^- \mu^+ + X$.
Parts a) and b) are for the Tevatron and LHC center of mass energies,
respectively.  The NLO (solid line) and LO (dashed line) cross sections are
shown. The cuts listed in Sec.~IIIB have been imposed.}
\label{FIG:ZZ:M}
\end{figure}

\begin{figure}
\caption{Same as Fig.~\protect{\ref{FIG:ZZ:M}} but for the inclusive
lepton transverse momentum distribution.
The 0-jet (dotted line) and 1-jet (dot-dashed line) exclusive cross sections
are also shown.}
\label{FIG:ZZ:PTL}
\end{figure}

\begin{figure}
\caption{Same as Fig.~\protect{\ref{FIG:ZZ:M}} but for the inclusive
lepton pseudorapidity distribution. }
\label{FIG:ZZ:YL}
\end{figure}

\begin{figure}
\caption{Same as Fig.~\protect{\ref{FIG:ZZ:M}} but for
the angular distribution of the $e^-$.  The angle $\theta_{e^-}$
is measured in the parent $Z$-boson rest frame with respect to
the parent $Z$-boson direction in the $ZZ$ rest frame.}
\label{FIG:ZZ:COSE}
\end{figure}

\begin{figure}
\caption{Same as Fig.~\protect{\ref{FIG:ZZ:M}} but for the angular
correlation between the lepton decay planes
[see Eq.~(\protect{\ref{EQ:CHI}})].}
\label{FIG:ZZ:COSCHI}
\end{figure}

\begin{figure}
\caption{The~transverse~cluster~mass~distribution~for~the~process
$p\,p\hskip-7pt\hbox{$^{^{(\!-\!)}}$} \to W^- W^+ + X
\to e^- \bar \nu_e \nu_e e^+ + X$.
Parts a) and b) are for the Tevatron and LHC center of mass energies,
respectively.  The NLO (solid line) and LO (dashed line) cross sections are
shown. The cuts listed in Sec.~IIIB have been imposed.}
\label{FIG:WW:CTM}
\end{figure}

\begin{figure}
\caption{Same as Fig.~\protect{\ref{FIG:WW:CTM}} but for the inclusive
charged lepton transverse momentum distribution.
The 0-jet (dotted line) and 1-jet (dot-dashed line) exclusive cross sections
are also shown.}
\label{FIG:WW:PTL}
\end{figure}

\begin{figure}
\caption{Same as Fig.~\protect{\ref{FIG:WW:CTM}} but for the
missing transverse momentum distribution. }
\label{FIG:WW:PTMISS}
\end{figure}

\begin{figure}
\caption{Same as Fig.~\protect{\ref{FIG:WW:CTM}} but for the inclusive
charged lepton pseudorapidity distribution. }
\label{FIG:WW:YL}
\end{figure}

\begin{figure}
\caption{The~transverse~cluster~mass~distribution~for~the~process
$p\,p\hskip-7pt\hbox{$^{^{(\!-\!)}}$} \to W^+ Z + X
\to \nu_e e^+ \mu^- \mu^+ + X$.
Parts a) and b) are for the Tevatron and LHC center of mass energies,
respectively.  The NLO (solid line) and LO (dashed line) cross sections are
shown. The cuts listed in Sec.~IIIB have been imposed.}
\label{FIG:WZ:CTM}
\end{figure}

\begin{figure}
\caption{Same as Fig.~\protect{\ref{FIG:WZ:CTM}} but for the inclusive
charged lepton transverse momentum distribution.
The 0-jet (dotted line) and 1-jet (dot-dashed line) exclusive cross sections
are also shown.}
\label{FIG:WZ:PTL}
\end{figure}

\begin{figure}
\caption{Same as Fig.~\protect{\ref{FIG:WZ:CTM}} but for the
missing transverse momentum distribution. }
\label{FIG:WZ:PTMISS}
\end{figure}

\begin{figure}
\caption{Same as Fig.~\protect{\ref{FIG:WZ:CTM}} but for the inclusive
charged lepton pseudorapidity distribution.}
\label{FIG:WZ:YL}
\end{figure}

\begin{figure}
\caption{Same as Fig.~\protect{\ref{FIG:WZ:CTM}} but for
the angular distribution of the $e^+$.  The angle $\theta_{e^+}$
is measured in the $W$-boson rest frame with respect to
the $W$-boson direction in the $WZ$ rest frame.}
\label{FIG:WZ:COSE}
\end{figure}

\begin{figure}
\caption{Same as Fig.~\protect{\ref{FIG:WZ:CTM}} but for
the angular distribution of the $\mu^-$.  The angle $\theta_{\mu^-}$
is measured in the $Z$-boson rest frame with respect to
the $Z$-boson direction in the $WZ$ rest frame.}
\label{FIG:WZ:COSMU}
\end{figure}

%
%

\begin{references}
\bibitem{CORRECTION}
In Refs.~\cite{ZZ}, \cite{WW}, and \cite{WZ}, the factor
$\lambda^{1/2}(\hat s,M^2_{V_1},M^2_{V_2})/(16 \hat s^2)$, where
$\lambda(a,b,c) = a^2 + b^2 + c^2 -2\,(ab + ac + bc)$,
was omitted from the virtual correction equations.
Also in Ref.~\cite{WW}, the last term in the equation defining $E_5$
in the appendix has the wrong sign; it should read $+\,8(U-T)/(S-4)$.
\end{references}
\end{document}